# Simulating the Power Electronics-Dominated Grid using Schwarz-Schur Complement based Hybrid Domain Decomposition Algorithm

Fatemeh Kalantari
University of Houston
Houston, USA
fkalantari@uh.edu

Jian Shi
University of Houston
Houston, USA
jshi23@central.uh.edu

Harish S. Krishnamoorthy
University of Houston
Houston, USA
hskrishn@central.uh.edu

*Abstract*— **This paper proposes a novel two-stage hybrid domain decomposition algorithm to speed up the dynamic simulations and the analysis of power systems that can be computationally demanding due to the high penetration of renewables. On the first level of the decomposition, a Schwarz-based strategy is used to decouple the original problem into various subsystems through boundary variable relaxation, while on the second level, each decoupled subsystem is further decomposed into subdomains that are solved independently using the Schur-complement approach. Convergence is checked in both stages to ensure that the parallelized implementation of the subsystems can produce identical results to the original problem. The proposed approach is tested on an IEEE 9 bus system in which one synchronous generator is replaced with a solar PV farm through a grid-forming inverter (GFM) with an admittance control method to evaluate its effectiveness and applicability for large-scale and very-large-scale implementations. Since conventional dual-loop GFMs are not stable when connecting to a stronger grid with a small grid inductance, a virtual inductance method is adopted to increase the equivalent inductance connecting the grid to enhance stability.**

## I. Introduction

Unlike traditional grid systems, smart modern green grids are made of much more active devices such as electric cars and many power sources like wind turbines and solar power with much more difficult behavior to predict, to name a few. In this paper, a novel two-stage hybrid domain decomposition algorithm is used to speed up the dynamic simulations and the analysis of such power systems with a high level of renewable penetration that can be computationally demanding since thousands of inverters operating at the same time create millions of scenarios that need to be handled precisely for a resilient and reliable future grid [1]. On the first level of the decomposition, the proposed algorithm incorporates a Schwarz-based strategy to decouple the renewable sources as well as synchronous machines (SM) from the network through boundary variable relaxation, and on the second level, each decoupled subsystem could be further decomposed into subdomains that are solvable independently using the Schur-complement approach. In this regard, a stable grid forming inverter with an admittance control method to be mathematically analyzed and modeled in MATLAB is selected so that Matpower can be used to simulate this low inertia system which demands load balance while stabilizing frequency by properly designing it. The two widely applied inverters in renewable energy generation systems are categorized as grid-following (GFL) inverters and grid-forming (GFM) inverters, however, GFLs face some challenges to provide inertia support. Additionally, the islanded operation is also a big challenge for GFLs which behave like current sources and rely on the grid voltage to realize synchronization and normal operation [2]. Differently, GFM inverters behave like voltage sources so that they have the islanded operation ability and can support the grid frequency to solve the frequency stability issues. Therefore, GFM inverters have attracted a lot of attention in recent years. A conventional dual-loop GFM inverter is prone to be unstable under a stronger grid condition because there is a small grid inductance connecting two voltage sources on grid-side and converter-side [3]. Hence, we decided to use a virtual inductance control method as in Fig. 1 which can increase the equivalent inductance to enhance stability to control GFM. Based on the eigenvalue analysis of these methods, its small-signal stability range is wider than the other inverters reported in the literature [4]. The closed-loop transfer functions of the current control loop can be represented as $G_i$, $L_f$ and $C_f$ are the filter inductance and capacitance, $L_c$ is the grid inductance and $v_g$ is the thevetin's equivalent circuit voltage of the power grid. Adding a virtual admittance $Y_v = 1 / (sL_v + R_v)$ as shown in Fig. 1. as the voltage control method can increase the equivalent inductance between two voltage sources to enhance the stability of the conventional dual-loop control method. The magnitude of the virtual admittance and the ratio of the virtual resistance and inductance $R_v/X_v$ are two key parameters to determine the small-signal stability of the GFM inverter system which will be addressed in the design section.

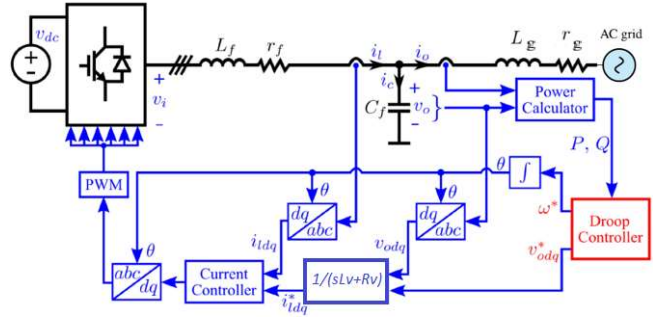

Fig. 1. Grid-forming inverter with a virtual admittance control method.

This paper is organized as follows. In Section II Algorithm formulation is described. Small signal equations of GFM are covered in Section III. Section IV presents the system dynamics and simulation results of our case study as well as numerical results followed by conclusion and future work in Section V.

## II. ALGORITHM FORMULATION

We have used a novel relaxation-based decomposition technique called Parallel-General-Norton with Multiple-port Equivalent (PGNME) technique which facilitates the distributed dynamic simulation and analysis for large-scale and very large-scale power systems with heavily polluted renewable penetration [5]. While previous work focus on traditional high inertia grid systems we have considered the low inertia modern systems including GFMs which have more complicated interaction with SMs and more complex dynamic calculations [7].

### A. Schwarz versus Schur-complement

There are two main categories defined in the literature for treating the interface values of a decomposed system, the first method is the Schur-complement method and the second one is Schwartz alternating methods. Schwarz decomposes the linear $AX=B$ equation while Schur-complement decomposes the system including the dynamic parts [8]. By defining the $Y$ matrix of the power system and selecting bus voltages as its variables, the Schur-complement methodology will be applied to parallelize the $YV=I$ equation which is an efficient way to precondition power system matrix $A$. LU factorization will be used to solve the linear system in MATLAB as the benchmark for a fare comparison.

### B. Subdomain interface variables processing

We consider a general system that is decomposed into various subsystems as shown in Fig. 2. It is assumed that within the many subsystems, subsystem $X$ is connected to subsystem $Y$ through a cut edge named $y_{XY}$. We denote the ports resulting from the cut edge, port $j$ for subsystem $X$ and port $k$ for subsystem $Y$. Now we can represent the boundary bus $j$ and $k$ as shown in Fig. 2 as well. Then, each subgraph can be considered as a linear passive network, where simulating subsystem $X$ requires solving the network equation as in (1):

$$V_{sub} = Y_{sub}{}^{-1}I_{sub} \qquad (1)$$

where $I_{sub}$ and $V_{sub}$ denote the node current and voltage within subsystem $X$ respectively and $Y_{sub}$ denotes the nodal admittance matrix. Then the boundary buses are added, $V_{dxj}$ is the boundary bus $j$ voltage and $S_{xj}$ is the boundary bus $j$ current injection [6]. $V_{dxk}$ is the boundary bus $k$ voltage and $S_{xk}$ is the boundary bus $k$ current injection. Now we must define the augmented admittance matrix with boundary buses included as $Y_{mod}$, then we can solve the subsystems in parallel as in (2):

$$V_{mod} = Y_{mod}{}^{-1}I_{mod} \qquad (2a)$$
$$V_{mod} = \begin{bmatrix} V_{sub}{}^T & V_{dx1} & V_{dx2} & ... & V_{dxn} \end{bmatrix}^T \quad (2b)$$
$$I_{mod} = \begin{bmatrix} I_{sub}{}^T & S_{x1} & S_{x2} & ... & S_{xn} \end{bmatrix}^T \qquad (2c)$$

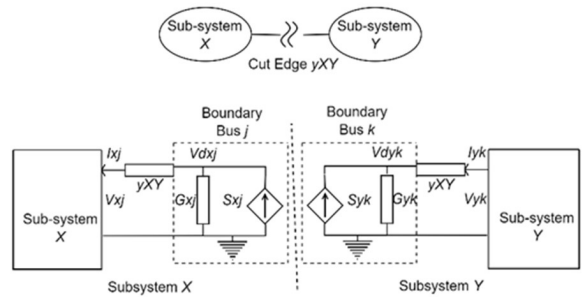

Fig. 2. Boundary bus creation and reconciliation.

Based on KVL and KCL, algebraic equations governing the interactions of subsystem $X$ and boundary bus $j$ are as in (3):

$$V_{d_{xj}} - V_{x_j} - I_{x_j}y_y{}^{-1} = 0 \qquad (3a)$$
$$G_{xj}V_{d_{xj}} = S_{x_j} - I_{xj} \qquad (3b)$$

These two constraints can be relaxed and expressed in the Jacobi-like iterative form for subsystems $X$ and $Y$ as in (4):

$$V_{d_{xj}}(l+1) = V_{yk}(l)I_{x_j}(l+1) = -I_{yk}(l) \qquad (4a)$$
$$V_{dyj}(l+1) = V_{xj}(l)I_{yk}(l+1) = -I_{xj}(l) \qquad (4b)$$

We obtain the corresponding parallel updating strategy for the controllable boundary state variables $S_{xj}$ and $S_{yk}$ as in (5):

$$S_{xj}(l+1) = G_{xj}V_{dxj}(l+1) + I_{xj}(l+1)$$
$$= G_{xj}V_{yk}(l) - I_{yk}(l) \qquad (5a)$$
$$S_{yk}(l+1) = G_{yk}V_{dyk}(l+1) + I_{yk}(l+1)$$
$$= G_{yk}V_{xj}(l) - I_{xj}(l) \qquad (5b)$$

This updating routine is repeated every relaxation iteration until the convergence criteria are met. The mismatch between the port currents is chosen as the convergence criteria as in (6)

$$\left| I_{xj}(l) + I_{yk}(l) \right| \leq \sigma \qquad (6)$$

where $\sigma$ is a very small number.

Schur-complement method for power system dynamic simulations is summarized as in (7):

$$\begin{pmatrix} D_1 & D_2 \\ D_3 & D_4 - \sum_i^{N-1} C_i S_i^{-1} B_i \end{pmatrix} \begin{pmatrix} \Delta V^{int} \\ \Delta V^{ext} \end{pmatrix} = \begin{pmatrix} g^{int} \\ g^{int} \sum_i^{M-1} C_i S_i^{-1} \tilde{f}_i \end{pmatrix} (7)$$

where $D_1$ represents the coupling between interior variables, $D_4$ represents the coupling between local interface variables, and $D_2$ and $D_3$ represent the coupling between the local interface and the interior variables [9]. $B_i$ and $C_i$ represent the coupling between the local interface variables and the external interface variables of the adjacent subdomains. $S_i$ is the local Schur-complement matrix and $f_i$ and g are the corresponding adjusted mismatch values.

## C.  *Two-stage Hybrid Decomposition Strategy*

The structure of the parallel code for the proposed two-stage hybrid decomposition algorithm can be described in Algorithm 1, depending on the scale of the power system, we can perform preconditioning in the first stage and adjust the number of subsystems to be solved in parallel.

| **Algorithm 1: The proposed two-stage hybrid domain decomposition algorithm** |
| --- |
| **Preconditioning and decomposition of subsystems** |
| 1:  Determine the number of subsystems and locations of cuts. |
| 2:  Apply Schwarz based PGNME to create each subsystem: |
| 3:  To find the multi-port equivalent of subsystems use (9). |
| 4:  To determine the diagonal and the off diagonal entries of the multi-port equivalent matrix use (11). |
| 5:  Calculate boundary bus parameters, $G$ and $S$ using (5). |
| 6:  Construct $Y_{mod}$ from *the Y* matrix for each subsystem, omitting the connection line at cut edges. |

| | |
|---|---|
| 7: | Initialize decomposed systems interface values as constants that make the subsystems decoupled. |
| 8: | Calculate initial machine states and check all steady states. |
| **Decomposition to subdomains using SC** | |
| 9: | Determine the number of subdomains and locations of cuts. |
| 10: | **While** the convergence criteria in (8) are met: |
| 11: | Update most recent subdomain currents as constants. |
| 12: | Apply SC algorithm for decomposition and solving subdomains using (11-12). |
| 13: | Relax $I_{xj}$ and $I_{yk}$ as in the Jacobi-like iterative form of (7). |
| 14: | Update boundary state variables $S_{xj}$ and $S_{yk}$ as in (8). |
| 15: | **End While** |

## III.    SMALL-SIGNAL EQUATIONS OF GFM

We have adopted the small-signal equations of GFM inverters with the admittance control method and have written the MATLAB code so that the model can be integrated into the two-level decomposition algorithm as a benchmark to be able to analyze any future large scale renewable polluted grids. To extract the state-space model of the virtual admittance method of a grid forming inverter circuitry we use the small-signal differential equations of the *Lf - Cf - Lg* circuits as shown in Fig. 1 in the system *d-q* frame given by (8)-(10) as derived in [10]:

$$
\begin{cases}
\Delta v_{Cd} - \Delta v_{gd} = Lg \dfrac{d\Delta i_{gd}}{dt} + R_g \Delta i_{gd} - \omega_1 L_g \Delta i_{gq} \\[2mm]
\Delta v_{Cq} - \Delta v_{gq} = Lg \dfrac{d\Delta i_{gq}}{dt} + R_g \Delta i_{gq} + \omega_1 L_g \Delta i_{gd}
\end{cases}
\quad (8)
$$

$$
\begin{cases}
\Delta i_{Ld} - \Delta i_{gd} = C_f \dfrac{d\Delta v_{Cd}}{dt} - \omega_1 C_f \Delta v_{Cq} \\[2mm]
\Delta i_{Lq} - \Delta i_{gq} = C_f \dfrac{d\Delta v_{Cq}}{dt} + \omega_1 C_f \Delta v_{Cd}
\end{cases}
\quad (9)
$$

$$
\begin{cases}
\Delta v_{cd}{}^{*} - \Delta v_{Cd} = L_f \dfrac{d\Delta i_{Ld}}{dt} + R_f \Delta i_{Ld} - \omega_1 L_f \Delta i_{Lq} \\[2mm]
\Delta v_{Cq}{}^{*} - \Delta v_{Cq} = L_f \dfrac{d\Delta i_{Lq}}{dt} + R_f \Delta i_{Lq} + \omega_1 L_f \Delta i_{Ld}
\end{cases}
\quad (10)
$$

Then the small-signal equations of the coordinate transformations according to the Park and iPark should be considered as (11)-(14), where $\Delta\theta_{ps}$ is the angle between the control *d-q* frame and the grid *d-q* frame [10]:

$$
\begin{cases}
\Delta v_{Cd}{}^{ctrl} = \Delta v_{Cd} + v_{Cq0} \cdot \Delta\theta_{ps} \\[2mm]
\Delta v_{Cq}{}^{ctrl} = \Delta v_{Cq} - v_{Cd0} \cdot \Delta\theta_{ps}
\end{cases}
\quad (11)
$$

$$
\begin{cases}
\Delta i_{gd}{}^{ctrl} = \Delta i_{gd} + i_{gq0} \cdot \Delta\theta_{ps} \\[2mm]
\Delta i_{gq}{}^{ctrl} = \Delta i_{gq} - i_{gd0} \cdot \Delta\theta_{ps}
\end{cases}
\quad (12)
$$

$$
\begin{cases}
\Delta i_{Ld}{}^{ctrl} = \Delta i_{Ld} + i_{Lq0} \cdot \Delta\theta_{ps} \\[2mm]
\Delta i_{Lq}{}^{ctrl} = \Delta i_{Lq} - i_{Ld0} \cdot \Delta\theta_{ps}
\end{cases}
\quad (13)
$$

$$
\begin{cases}
\Delta v_{cd}{}^{*} = \Delta v_{cd}{}^{*\,ctrl} - v_{cq0} \cdot \Delta\theta_{ps} \\[2mm]
\Delta v_{cq}{}^{*} = \Delta v_{cq}{}^{*\,ctrl} + v_{cd0} \cdot \Delta\theta_{ps}
\end{cases}
\quad (14)
$$

Besides, the small-signal linearized equations of the current control loops are as in (15) and (16):

$$\begin{cases} \dfrac{d\Delta Int_{id}}{dt} = \Delta i_{Ld}{}^* - \Delta i_{Ld}{}^{ctrl} \\ \Delta v_{cd}{}^{*ctrl} = K_{p-id}\Big(\Delta i_{Ld}{}^* - \Delta i_{Ld}{}^{ctrl}\Big) + K_{i-id}\Delta Int_{id} - \omega_1 L_f \Delta i_{Lq}{}^{ctrl} \end{cases}$$

(15)

$$\begin{cases} \dfrac{d\Delta Int_{iq}}{dt} = \Delta i_{Lq}{}^* - \Delta i_{Lq}{}^{ctrl} \\ \Delta v_{cq}{}^{*ctrl} = K_{p-iq}\Big(\Delta i_{Lq}{}^* - \Delta i_{Lq}{}^{ctrl}\Big) + K_{i-iq}\Delta Int_{iq} - \omega_1 L_f \Delta i_{Ld}{}^{ctrl} \end{cases}$$

(16)

and the small-signal linearized equations of the active power and reactive power are as in (17) and (18):

$$\Delta P^{ctrl} = \frac{3}{2}\begin{bmatrix} i_{gd0} & i_{gq0} \end{bmatrix} \cdot \begin{bmatrix} \Delta v_{Cd}{}^{ctrl} \\ \Delta v_{Cq}{}^{ctrl} \end{bmatrix} + \frac{3}{2}\begin{bmatrix} v_{Cd0} & v_{Cq0} \end{bmatrix} \cdot \begin{bmatrix} \Delta i_{gd}{}^{ctrl} \\ \Delta i_{gq}{}^{ctrl} \end{bmatrix}$$

(17)

$$\Delta Q^{ctrl} = \frac{3}{2}\begin{bmatrix} -i_{gq0} & i_{gd0} \end{bmatrix} \cdot \begin{bmatrix} \Delta v_{Cd}{}^{ctrl} \\ \Delta v_{Cq}{}^{ctrl} \end{bmatrix} + \frac{3}{2}\begin{bmatrix} v_{Cq0} & v_{Cd0} \end{bmatrix} \cdot \begin{bmatrix} \Delta i_{gd}{}^{ctrl} \\ \Delta i_{gq}{}^{ctrl} \end{bmatrix}$$

(18)

In the next step, the small-signal differential equations of the first-order low-pass filter (LPF) are obtained where $\omega_{LPF}$ is the cut-off angular frequency of the LPFs as in (19):

$$\begin{cases} \dfrac{d\Delta P^{CTRL}{}_{LPF}}{dt} + \omega_{LPF}\Delta P^{ctrl}{}_{LPF} = \omega_{LPF}\Delta P^{ctrl} \\ \dfrac{d\Delta Q^{CTRL}{}_{LPF}}{dt} + \omega_{LPF}\Delta Q^{ctrl}{}_{LPF} = \omega_{LPF}\Delta Q^{ctrl} \end{cases}$$

(19)

Furthermore, the small-signal equations of the $P$-droop and $Q$-droop control are as in (20):

$$\begin{cases} \dfrac{d\Delta \theta_{ps}}{dt} = \Delta \omega_{ps} = m_p\Big(\Delta P^* - \Delta P^{ctrl}{}_{LPF}\Big) \\ \Delta E = n_q\Big(\Delta Q^* - \Delta Q^{ctrl}{}_{LPF}\Big) \end{cases}$$

(20)

Regarding the virtual admittance method in Fig. 1, the small-signal equations of current reference values are as in (21):

$$\begin{cases} L_v\dfrac{d\Delta i_{Ld}{}^*}{dt} + R_v\Delta i_{Ld}{}^* - \omega_1 L_v \Delta i_{Lq}{}^* = \Delta E - \Delta v_{Cd}{}^{ctrl} \\ L_v\dfrac{d\Delta i_{Lq}{}^*}{dt} + R_v\Delta i_{Lq}{}^* + \omega_1 L_v \Delta i_{Ld}{}^* = 0 - \Delta v_{Cq}{}^{ctrl} \end{cases}$$

(21)

According to (8)-(21), the state-space model of a grid forming inverter with virtual admittance control method can be derived as (22), which are described by thirteen linear first-order differential equations.

$$\Delta \dot{x}_{3(13\times1)} = A_{3(13\times13)} \cdot \Delta x_{3(13\times1)} + B_{3(13\times4)} \cdot \Delta u_{3(4\times1)} \quad (22)$$

where $\Delta x_{3(13\times1)} = [\Delta i_{gd}, \Delta i_{gq}, \Delta v_{Cd}, \Delta v_{Cq}, \Delta i_{Ld}, \Delta i_{Lq}, \Delta Int_{id}, \Delta Int_{iq}, \Delta i_{Ld}^*, \Delta i_{Lq}^*, \Delta P_{LPFctrl}, \Delta \theta_{ps}, \Delta Q_{LPFctrl}]^T$ and. $\Delta u_{3(4\times1)} = [\Delta v_{gd}, \Delta v_{gq}, \Delta P^*, \Delta Q^*]$. Also, $i_{gd}$ and $v_{gd}$ are the direct current and voltage components while $i_{gq}$ and $v_{gq}$ are the quadrature current and voltage components of the grid. $v_{Cd}$ and $v_{Cq}$ refer to the capacitor filter voltage in the $d$-$q$ frame and $\Delta \theta_{ps}$ is the angle between the control $d$- $q$ frame and the grid $d$-$q$ frame. Considering the initial values of the variables, we used three different MATLAB functions of Modified Euler, Runge Kutta-Fehlberg, and Runge Kutta-Higham to solve the circuit. The stability depends on $A$, and the matrixes $A_{3(13\times13)}$ and $B_{3(13\times4)}$ are given in [11]. Since the power has a weak impact on the small-signal stability, the zero-power condition is used for analysis in this paper. The system and control parameters of a 30 kW GFM inverter are used. When the virtual admittance is designed properly, the inverter is stable no matter if the grid is strong or weak. We will use a multi-converter system consisting of identical modules to be rated at 100 MVA, which is equal to the SM rating of the modified IEEE 9 bus system. Each module is interfaced to a medium voltage line via an LV-MV transformer [12].

## IV. CASE STUDY

The proposed approach is tested on a modified IEEE 9 bus system to evaluate its effectiveness. We keep two traditional synchronous generators (SMs) at buses two and three and replace the first SM at bus one with a GFM inverter connected to a PV farm as shown in Fig. 3. As directed by the first stage of algorithm described in the previous section GFM is considered as the first domain and the rest of the system is categorized in the second stage. The strategy is to divide the system in a way that the mathematical workload is divided fairly among the cores so that we would be able to apply our algorithm on a parallel hardware such as GPUs. In the second stage of the algorithm, we have divided the rest of the network equally to subdomain 1 and subdomain 2 as shown in Fig. 3.

Using Simulink, we replace the virtual admittance control block in Fig. 1 with a PD block where $R_v$ replaces for the proportional constant, P, while $L_v$ replaces for the differential constant, D, and then we use the Reciprocal block to get the exact equation of $(1/(Rv+sLv))$.

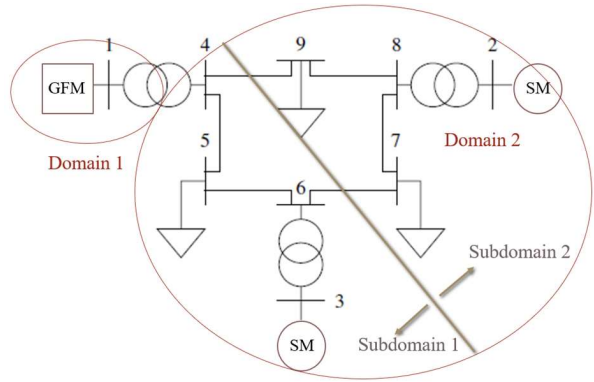

Fig. 3. Modified IEEE 9 bus test system with two synchronous machine, and one large-scale multi-converter GFM.

Remark 1. In this case study, the grid forming inverter in Figure 3 is an aggregate model of commercial converter modules with parameters described in Table I in details. In order to replace the synchronous generator of the original IEEE9bus system with the aggregate GFM model we need to come up with the same rating of 100 MVA of the SM. In this regard 200 converter modules rated at 500 kVA are used which should be interfaced to a medium voltage line via LV-MV transformers. We have connected 100 commercial transformers in parallel rated at 1:6 MVA to do the job as can be seen in Fig. 4. The detailed presentation and derivation of the model aggregation is developed analogous to [15]–[17].

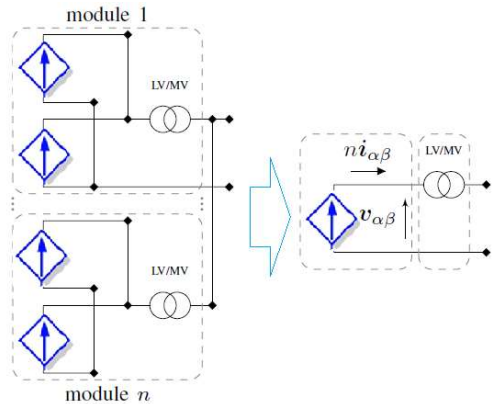

Fig. 4. Equivalent model of large-scale multi-converter system consisting of $n$ identical modules (left) and aggregate model (right).

## A. System Dynamics

To address dynamic parts of the system with accurate details, the MatDyn package is used. Regarding transient stability simulation, the transmission network is modeled in the main frequency phasor domain, and the dynamics of the system depend not only on rotating machines and control devices such as excitation system, power system stabilizer, turbine, and governor but also on the required fast power electronic devices such as grid forming inverters to merge renewable sources to the grid. We should emphasize that in the mixed SM-GFM system, it is vital to account for dc side limits (either through AC current limits or dc voltage measurements) to stabilize the GFCs in presence of SMs. Both the 4$^{th}$ order model and classical generator model are used in the dynamic simulation development. the equations of motion for generator $i$ are as in (23):

$$\frac{2H_i}{\omega_s} \cdot \frac{d\omega_i}{dt} = P_{mi} - P_{ei} - D_i\left(\omega_i - \omega_s\right), \quad \frac{d\delta_i}{dt} = \omega_i - \omega_s \quad (23)$$

where $H_i$ is the inertia constant of generator $i$ normalized by the system base, $\omega_i$ is the speed for generator $i$, $\omega_s$ is the synchronous speed, $P_m$ and $P_e$ are the mechanical power input and active power at the air gap of generator $i$, $D_i$ is the damping coefficient, and $\delta_i$ is the angular position of the rotor of generator $i$ in the electrical radians concerning a synchronously rotating reference [13]. Moreover, the load is represented by the passive impedance. For the generator exciter, model 2 IEEE DC1A excitation system is used. The most stable grid forming inverter reported in the literature with the admittance controllability method is adapted and modeled mathematically in MATLAB to connect a solar farm to the grid. The parameters used in the implementation are reported in Table I [14] including IEEE 9-bus test system base values, synchronous machine (SM) parameters, single converter module parameters used in ODE equations as well as ac current control and droop control loop parameters.

TABLE I: Case study model and control parameters [14].

| IEEE 9-bus test system base values | | | | | |
|---|---|---|---|---|---|
| $S_b$ | 100 MVA | $v_b$ | 230 kv | $\omega_1$ | 2π60 rad/s |
| **Synchronous machine (SM)** | | | | | |
| $S_r$ | 100 MVA | $v_r$ | 13.8 kv | $D_f$ | 0 |
| $H$ | 3.7 s | $dp$ | 1% | $\tau_g$ | 5 s |
| **Single converter module** | | | | | |
| $R_f$ | 0.001 Ω | $L_f$ | 5 mH | $C_f$ | 10 µf |
| $n$ | 100 | $\tau_{dc}$ | 50 ms | $I_{dc\,max}$ | 1.1 pu |
| $R_g$ | 57 m Ω | $L_g$ | 15 mH | $S_r$ | 500 KVA |
| $R_v/X_v$ | 0.2 | $L_v$ | 30 mH | $\omega_{1LPF}$ | 300 rad/s |
| **Droop control** | | | | | |
| $m_p$ | 0.25$\omega_1/S_b$ | $n_q$ | 0.25$v_b/S_b$ | $\omega_1$ | 2π60 rad/s |
| **AC Current control** | | | | | |
| $K_{p\_id}$ | 5 | $K_{i\_id}$ | 5000 | $K_{p\_iq}$,$K_{i\_iq}$ | 0 |

## B. System Simulation Results

In this paper which reports on preliminary tests, our proposed two stage algorithm has divided the modified IEEE 9 bus system into two subsystems. In the first stage we apply Schwarz based PGNME method to decompose the solar PV farm which is connected to the grid through 200 commercial GFM converters and 100 transformers from the system, and in the second stage of decomposition, the remaining system is decomposed into two subdomains to be solved in parallel for a more efficient faster solution as shown in Fig. 3. All values are in per unit on a 100-MVA base with a frequency of 60 Hz, the step size was set to 0.01 second in the half-second after the initial disturbance and then increased to 0.05 second. The terminal voltage and current of the grid forming inverter are obtained using MatDyn as the initial conditions to the first order differential equations to solve the thirteen ODE after modeling the grid forming inverter. We deployed MATLAB internal ODE solvers such as Modified Euler and Runge Kuttato confirm that results match perfectly to the benchmark as shown in Fig. 4. We consider a bus-bar short circuit fault occurring at bus 2 at time t = 1.2 seconds and clearing in 0.2 seconds later at 1.4 seconds as the first scenario in analysis. The fault will affect all the voltage buses at the instant of occurrence, as we can observe in Fig. 5. The dashed lines indicate the results of the benchmark for voltage at buses one and two which matches the results of the hybrid two-stage algorithm for the same buses after decomposition accordingly. In the second scenario, the sixth line admittance is changed during simulation by defining the "line-change" matrix at t = 1.5 seconds. The algorithm converged after 2 iterations. Another observation is the fast response of grid forming converter compared to the slow turbine dynamics, where GFC reaches frequency synchronization at a faster timescale and then synchronizes with the SM. Overall, for any given disturbance input, GFMs can react faster than SM [11]. The simulation results confirmed that after a 0.75 pu load increase, the converters quickly synchronized with the system frequency and then slowly synchronized with the machine.

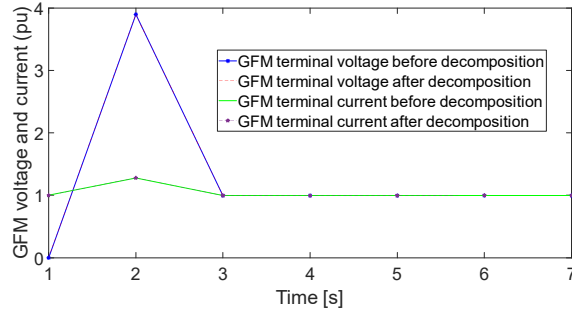

Fig. 4. Enhanced Admittance GFM terminal voltage and current in pu.

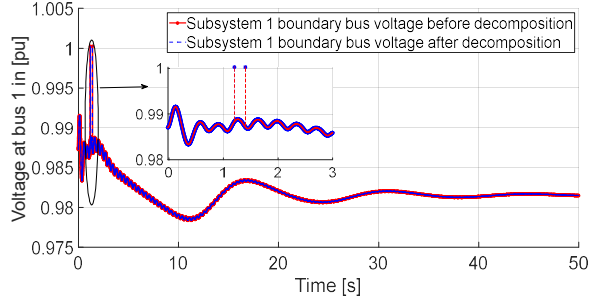

Fig. 5. Boundary bus 1 voltages for subsystem 1 before and after decomposition in the first scenario.

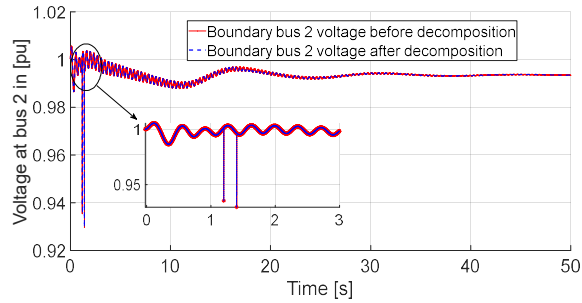

Fig. 6. Boundary bus 2 voltages for subsystem 2 before and after decomposition in the first scenario.

## V. Conclusion and Future Work

In this study the two-stage hybrid algorithm decomposed the modified IEEE 9 bus power system into two subsystems and two subdomains, converging after only 3 iterations where a smaller number of Jacobian updating and factorizations are needed which will lead to speeding up the whole system especially if we are facing a large-scale power system with thousands of DAE equations to be addressed due to the high number of power electronic devices engaged in the future highly polluted renewable grids confirming the importance of this fast accurate algorithm technique. The grid forming inverter was mathematically modeled in MATLAB using thirteen first-order differential equations to be submerged to bus one of the IEEE 9 bus system replacing its traditional generator which was decomposed in the first level for a parallel solution. In all cases, the simulation results due to the deployment of the two-stage hybrid algorithm and LU factorization benchmark matched accordingly. As next step, the proposed algorithm will be deployed for large-scale transient stability simulation of electrical power systems on graphics processing units (GPUs) which have shown great potential for calculating large-scale numerical problems. Also, we will consider the proposed algorithm to be deployed for large-scale transient stability simulation of ships' electrical power systems including the DC-DC converters.